# Visualization of multifractal superconductivity in a two-dimensional transition metal dichalcogenide in the weak-disorder regime

Carmen Rubio-Verdú[1], Antonio M. García-García[*,2], Hyejin Ryu[3,4], Deung-Jang Choi[1,5,6,7], Javier Zaldívar[1], Shujie Tang[3,8], Bo Fan[2], Zhi-Xun Shen[8,9], Sung-Kwan Mo[3], José Ignacio Pascual[1,7] and Miguel M. Ugeda[*,1,5,6,7]

[1]CIC nanoGUNE, 20018 Donostia-San Sebastián, Spain.

[2]Shanghai Center for Complex Physics, Department of Physics and Astronomy, Shanghai Jiao Tong University, Shanghai 200240, China.

[3]Advanced Light Source, Lawrence Berkeley National Laboratory, Berkeley, CA 94720, USA.

[4]Center for Spintronics, Korean Institute of Science and Technology, Seoul 02792, Korea.

[5]Centro de Física de Materiales CFM/MPC (CSIC-UPV/EHU), 20018 San Sebastián, Spain.

[6]Donostia International Physics Center (DIPC), 20018 San Sebastián, Spain.

[7]Ikerbasque, Basque Foundation for Science, 48013 Bilbao, Spain.

[8]Stanford Institute for Materials and Energy Sciences, SLAC National Accelerator Laboratory, Menlo Park, CA 94025, USA.

[9]Geballe Laboratory for Advanced Materials, Departments of Physics and Applied Physics, Stanford University, Stanford, CA 94305, USA.

\* Corresponding authors: amgg@sjtu.edu.cn and mmugeda@dipc.org

## Abstract

Eigenstate multifractality is a distinctive feature of non-interacting disordered metals close to a metal-insulator transition, whose properties are expected to extend to superconductivity. While multifractality in three dimensions (3D) only develops near the critical point for specific strong-disorder strengths, multifractality in 2D systems is expected to be observable even for weak disorder. Here we provide evidence for multifractal features in the superconducting state of an intrinsic weakly disordered single-layer $NbSe_2$ by means of low-temperature scanning tunneling microscopy/spectroscopy. The superconducting gap, characterized by its width, depth and coherence peaks' amplitude, shows a characteristic spatial modulation coincident with the periodicity of the quasiparticle interference pattern. Spatial inhomogeneity of the superconducting gap width, proportional to the local order parameter in the weak-disorder regime, follows a log-normal statistical distribution as well as a power-law decay of the two-point correlation function, in agreement with our theoretical model. Furthermore, the experimental singularity spectrum $f(\alpha)$ shows anomalous scaling behavior typical from 2D weakly disordered systems.

## Introduction

Quantum coherence phenomena have a profound impact in the dynamics of disordered media. A paradigmatic example is the Anderson transition in disordered metals where quantum interference of non-interacting electrons induces spatial localization, leading to insulating states beyond a critical disorder strength (*1*). In the vicinity of the transition preceding localization, these systems display eigenstates being neither extended nor localized that strongly fluctuate at all length scales. Such near-critical eigenstates exhibit multifractal character, i.e., they are formed by interwoven sets of different fractals, each characterized by a non-integer dimension(*2–4*). These critical eigenstates' correlations are of fundamental relevance in the presence of disorder since the multifractal regime dominates their electronic and magnetic properties(*5–8*).

When attractive interactions between electrons are present in metals, coherent electronic states such as superconductivity (SC) can emerge in the presence of disorder. In the strong-disorder regime beyond the critical value, Anderson localization disables long-range quantum coherence, thus quenching superconductivity. For weak-disorder, superconductivity persists (*9*, *10*) even in polycrystalline or amorphous materials near the Anderson localization transition(*11*). Nonetheless, even weak disorder strongly affects superconductivity. Recent experimental studies showed that disorder leads to spatial inhomogeneity(*10*, *12–14*) and granularity(*15*) in the superconducting order parameter, in agreement with previous theoretical results(*16*). Despite these findings demonstrate the intricate interplay between disorder and superconductivity, the existence of the superconducting state in the multifractal regime remains unexplored, and the signatures of multifractality have been mostly theoretically addressed so far(*17–20*). Such investigation seems particularly suitable in 2D materials since, unlike in 3D where multifractality only arises in a narrow range of disorder around criticality, multifractality is expected to emerge in 2D for a broad range of disorder strengths(*20*, *21*). The existence of multifractality in 2D superconductors is expected to shed light on long standing problems such as the observed intermediate metallic state(*22*), and variations of the superconductivity strength near the 2D limit (*23*).

2D is the marginal dimension for both localization and superconductivity. Scaling theory predicts that electronic eigenstates in infinite 2D systems are localized regardless of the disorder strength(*24*) precluding the development of multifractality and superconductivity. However, this is only valid for infinite 2D systems with time reversal symmetry and, therefore, 2D systems with spin-dependent hopping exhibit a metallic ground state for sufficiently weak disorder and, therefore, a broad region where both multifractality and superconductivity can coexist (*25*). 2D materials develop multifractality provided the size of the material is smaller than the localization length η, which can be extremely large in the weak-disorder regime (it scales inversely with the disorder strength, $\eta \propto l \cdot \exp(\pi k \cdot l/2)$, where $k$ is the wavevector and $l$ the mean-free path). In this arena, single-layer transition metal dichalcogenide superconductors are envisaged as ideal systems to investigate this elusive regime.

Here we provide evidence for the multifractal character of the superconducting state in single-layer $NbSe_2$, a 2D superconductor(*23*). Intrinsic weak-disorder in $NbSe_2$ monolayers triggers multifractality of the single-particle eigenstates, which dramatically impacts its superconducting state. By means of spatially-resolved scanning tunneling spectroscopy at T = 1.1 K, well below $T_C$, we observe strong sub-nm-sized fluctuations in the superconducting order parameter (proportional to the SC width for weak-disorder(*16*)) as well as in the coherence-peak amplitude. We find that the spatial distribution of the SC order parameter amplitude corresponds to a log-normal type and, simultaneously, its spatial correlations show power-law decay for intermediate distances. Furthermore, the associated experimental singularity spectrum *f(α)* shows anomalous scaling behavior typical from 2D weakly disordered systems. These features demonstrate that superconductivity in single-layer $NbSe_2$ is governed by multifractal electronic states even for weak disorder.

**Results**

Our experiments were performed on single-layer NbSe$_2$ grown on bilayer graphene (BLG)/6H-SiC(0001) as shown in Fig. 1a. Atomically-resolved STM images of the NbSe$_2$ films (Fig. 1b) reveal high crystallinity ($< 1 \cdot 10^{12}$ defects/cm$^2$), where the main source of defects are island edges and 1D grain boundaries (Fig. 3b). Superconductivity in NbSe$_2$ is depressed in the single-layer limit (T$_C$ = 1.9 K)(*23*) as compared to the bulk (T$_C$ = 7.2 K). At T = 1 K, charge density wave (CDW) order is fully developed as seen in Fig. 1b. STM dI/dV spectra (Fig. 1c) taken at different locations of the same region exhibit a dip in the density of states (DOS) at the Fermi level (E$_F$) that corresponds to the superconducting gap. The features that define the SC gap, i.e. depth, width and coherence peaks amplitude, are seen to locally vary at the nm-scale. These fluctuations were consistently observed in all the NbSe$_2$ regions studied regardless of shape, size and crystallinity.

To better understand the nature of the SC fluctuations in single-layer NbSe$_2$, we spatially mapped the width and depth of the SC gap in multiple regions with high-spatial resolution of ~1 Å (see supplementary information (SI) for the extraction procedure of the SC gap). The width and depth of the SC gap are a measure of the local SC order parameter amplitude for weak-disorder(*16*) and the degree of development of SC, respectively. Figure 2a shows a representative map of the spatial distribution of the width of the SC gap in a 12.4 nm x 12.4 nm region (90 x 90 mesh that yields 8100 values). The width map unveils clear spatial fluctuations of the SC order parameter in single-layer NbSe$_2$ within the nm-scale. Fourier transform (FFT) analysis of this width map (Fig. 2b) yields a reciprocal-space ring of radius q = 0.88 ± 0.08 Å$^{-1}$, which reveals a single wavelength of λ = 7 ± 1 Å involved in the complex pattern of the SC fluctuations in real-space. The depth of the SC gap exhibits similar spatial fluctuations with the same wavelength (Fig. 2c). These nanoscale spatial variations λ are smaller than the SC coherence length for single-layer NbSe$_2$ (ξ(0) ~10 nm)(*26, 27*), which we emphasize that is theoretically plausible (see SI for an extended discussion) and was also observed in several superconductors with different dimensionalities(*10, 28–30*).

To confirm the superconducting inhomogeneity in single-layer NbSe$_2$, we compare the spatial fluctuations of the order parameter (Fig. 2a) with the spatial distribution of the amplitude of the coherence peaks (Fig. 2d) acquired over the same region. The peaks' amplitude is intimately related to long-range superconducting phase coherence and directly proportional to the quasiparticles' lifetime, which can be reduced by multiple mechanisms present in reduced dimensions and by the presence of disorder(*10, 11, 13, 31*). The maps show strong peak's amplitude fluctuations over the same length scale as that seen for the SC width and depth (7 Å), although surrounded by regions where the coherence peaks are depleted due to intrinsic disorder in the 2D superconductor. Despite these fluctuations in the phase-coherence at the submicron-scale, mesoscopic transport measurements in this kind of samples revealed that phase coherence holds(*23*).

Herein we focus on the origin of the characteristic wavelength of the SC fluctuations of ~ 7 Å. Such wavelength does not match either the atomic lattice (3.44 Å), the SiC reconstruction (32 Å) or the CDW superlattice (10.3 Å). The latter is expected since the CDW opens only on a specific spots on the Nb K-H sheets, leaving most of the Fermi surface to superconductivity (*32*). In order to reveal its origin, we performed spatially-resolved dI/dV mapping of the electronic structure of single-layer NbSe$_2$ near E$_F$. Figure 3a shows a typical conductance map (dI/dV(**r**, E)) at V$_b$ = + 40 mV in a defective region (corresponding topography in Fig. 3b). Defects act as scatterers giving rise to quasiparticle interference (QPI) patterns, extending tens of nanometers away from them. Figure 3c shows the Fourier analysis of the QPI map of Fig. 3a, where an anisotropic ring with intensity maxima (**q**$_{qpi}$) is present along the same ΓM direction as the dispersionless CDW signal (**q**$_{cdw}$ ≅ ΓM/3). This feature is observable within ± 100 meV and shows a slight dispersion in *k* along ΓM (fig. 3d). The wavelength of the QPI patterns is λ$_{QPI}$ = 6.4 ± 0.2 Å, which nearly matches the periodicity of the SC fluctuations. This suggests that the SC fluctuations and the QPI-induced charge modulations likely share a common origin. These QPI patterns, previously observed in bulk 2H-NbSe$_2$, are attributed to enhanced backscattering due to strong direction-dependent electron-phonon interactions(*33*). *Ab-initio* calculations indicate that soft acoustic phonons along the ΓM direction are strongly coupled to electrons(*34*). This is a plausible origin of the spatial SC modulations given the significant role of acoustic phonons in the Cooper pairs formation. A correlation between the SC modulations and DOS fluctuations, i.e., Δ(r) ∝ LDOS can be ruled out due the highly dynamic conductance within the SC gap (see SI).

**Discussion**

Statistical analysis of the SC width values from the spatially resolved maps reveals relevant features of the SC fluctuations related to multifractality. Figure 4a (upper plot, in orange) shows the probability distribution of the SC gap width for the map shown in fig. 2a. The values fluctuate over a wide energy range of 0.8 meV around a mean value of 1.1 meV. Such a broad distribution reflects the large amplitude of the fluctuations of the SC order parameter induced in the weak-disorder (intrinsic) regime(*16*). Three additional experimental distributions of the SC width from regions with different degrees of intrinsic disorder are shown in figs. 4a (log scale) and 4b (linear scale). All of them exhibit a marked right-skewed behavior with different degrees of asymmetry. The larger the width of the distributions, the larger their asymmetry, presumably due to stronger local disorder. This marked skewness is only observed in the statistics of the SC order parameter (SC width), being the distributions of the SC gap depth and coherence peaks nearly symmetric and Poisson-type, respectively (SI).

In order to understand the characteristic properties of the SC gap distributions, we modelled single-layer NbSe$_2$ as a system close to the Anderson metal-insulator transition, where electronic states exhibit a multifractal character, i.e. anomalous scaling of the inverse participation ratio and a power-law decay of eigenstate correlation functions(*2*, *3*, *35*) (SI). In 3D, where multifractality only emerges in the strong disordered limit ($\gamma = g^{-1} \sim 1$, with g the Thouless conductance), mean field theory found that multifractality enhances the superconducting $T_C$(*17*, *18*). In 2D, where multifractality occurs in the weak-disordered regime ($\gamma \ll 1$), similar results were obtained(*19*). A more realistic theoretical analysis(*20*) of the spatial structure of the SC gap in 2D for $\gamma \ll 1$ found that, in agreement with previous experimental results (*25*, *36*), $T_C$ was enhanced for sufficiently weak coupling, though more modestly than in earlier predictions(*16*, *17*). The SC gap spatial distribution is modelled by a log-normal distribution developed in (*18*):

$$P\left(\frac{\Delta(r)}{\bar{\Delta}}\right) = \frac{\bar{\Delta}}{\Delta(r)\sqrt{2\pi\gamma\ln(E_0/\epsilon_D)}} exp\left[-\frac{\left(ln\left(\frac{\Delta(r)}{\bar{\Delta}}\right) - \frac{3}{2}\gamma\ln(\epsilon_D/E_0)\right)^2}{2\gamma\ln(E_0/\epsilon_D)}\right] \quad \text{(eq. 1)}$$

with $\epsilon_D$ the Debye energy, $E_0$ the minimum energy scale to observe multifractal eigenvector correlations, and $\bar{\Delta}$ is the average gap (see SI). Here $\gamma$ is proportional to the disorder strength and eq. 1 is valid for $\gamma = 4/k_F \cdot l \ll 1$ (see materials and methods). We use this expression to fit the statistical distributions of the SC width maps measured in different sample regions. The theoretical SC gap $\Delta(r)$ corresponds(*16*) to the experimental SC width in the weak disorder limit ($\gamma \ll 1$) of interest for this work. As shown in Fig. 4a, the experimental right-skewed distributions of the SC width are in all cases better described by a log-normal distribution in the multifractal regime (colored dashed lines) than by a Gaussian distribution (grey dashed lines). These fits yield in all cases small $\gamma$ values of $\approx 0.1$ (values in fig. 4a), which indicates that the explored NbSe$_2$ regions are barely defective.

A natural question that arises is whether other mechanisms could lead to log-normal distributions and whether this feature obtained from a BCS approach occurs in more realistic theoretical frameworks. First, we can rule out thermal effects since the temperature dependence of the SC gap is rather weak even beyond $T_C/2 = 1$ K. Furthermore, an indication of thermal effects would be the development of a peak in the distribution at $\Delta = 0$ corresponding to locations where the SC vanishes, which we do not observe. Thermal fluctuations beyond the employed mean field formalism are also relevant only close to $T_C$. We can also rule out quantum fluctuations since they are suppressed in the weak-disordered limit ($\gamma \ll 1$). Furthermore, our recent calculations using the Bogoliubov de Gennes (BdG) formalism reproduce the SC gap log-normal distribution (not shown here).

To corroborate the multifractal character of the superconducting state in single-layer NbSe$_2$, we investigate the spatial correlations of the SC gap $\Delta(r)$, a fundamental property of the multifractal state. In addition to the log-normal distribution, another signature of multifractality is the power-law decay of eigenstate correlations for length scales larger than the mean-free path ($l$)(*37*), which we estimate here to be of $\sim 2$ nm (see materials and methods). Fig. 4c shows the two-point spatial correlation function of the SC order parameter from the SC gap width map of Fig. 2a, which is directly related to the two-point correlation of multifractal eigenstates. The observed decay of the correlations can be fitted to a power-law function restricted to intermediate distances as (see SI):

$$<\Delta(r)\Delta(r')> \propto 1/|r-r'|^\gamma \quad \text{(eq. 2)}$$

with the same exponent γ that governs the decay of the multifractal eigenstates (γ = 0.2 ± 0.1 in Fig. 4c). The power-law decay is a feature present in all the studied regions for relatively long scales ($0.7\ nm < |r-r'| < 7\ nm$). The fits to power-law functions in the studied regions yield γ values that are in qualitative agreement with those independently obtained from the log-normal distributions. In agreement with our predictions(*20*), islands with smaller γ values present a power-law decay followed by faster decay (likely exponential) for longer distances.

Lastly, the SC order parameter is expected to show multifractal features as is built up from multifractal eigenstates of the one-body problem. The singularity spectrum *f(α)* is an observable of a multifractal measure that can be computed (see SI), and it represents the ensemble of scaling dimensions that characterize the multifractal entity. In figs. 5a-d we plot *f(α)* for the previously shown experimental Δ(r) distributions. In all the cases, the shape of *f(α)* is well described by a parabola, typical of multifractal eigenstates in weakly disordered 2D systems(*35*, *38*), which becomes broader as disorder (represented by γ) increases suggesting an anomalous scaling in the spatial distribution of Δ(r). In the weak disorder limit at criticality, *f(α)* is exactly parabolic in the non-interacting limit and can be fitted (straight lines in figs. 5a-d) as in refs. (*35*, *38*). This experimental behavior, which confirms the SC order parameter as multifractal, is qualitatively reproduced by numerically computing *f(α)* for a 2D system in a random potential using the BdG equations (see SI).

These unique features observed in the SC state of single-layer $NbSe_2$ must be originally triggered by the multifractal structure of the electronic states (strictly the only fractal entity in the system) (*17–20*). We have explored their nature by simultaneously mapping the conductance (LDOS) in the same regions where the SC gap was characterized. Interestingly, multifractal effects on the electronic conductance are only observable in the most disordered regions (SI, fig. S2m-p). The conductance distributions corresponding to the four regions studied in Fig. 4 show Gaussian distributions for γ ≤ 0.16 that evolve towards a right-skewed log-normal distribution for larger disorder (γ = 0.142). The larger sensitivity of the SC to multifractality than that of the LDOS is due to the SC gap exponential dependence on the electron-phonon coupling *(V)*, i.e., $\Delta \propto e^{-1/N(0)V}$, with *N(0)* the LDOS at $E_F$. Small changes in the coupling and DOS induce comparatively large changes in the SC gap for disorder regimes where the electronic states are barely multifractal and the effects in the conductance nearly imperceptible. Therefore, superconductivity amplifies the effect of multifractality and enables its observation in nearly pristine 2D superconductors.

In summary, we provide experimental evidence of the elusive multifractal superconducting state in a prototypical 2D superconductor in the low-disorder (intrinsic) regime. We demonstrate that multifractal characteristics fully manifest in the superconducting state even in the weak disorder regime. Multifractality is therefore expected to dominate the superconducting properties of the recently discovered family of highly crystalline 2D superconductors with spin-orbit coupling such as single layers of transition metal dichalcogenides. These 2D materials open the door for further investigation and eventual control of the intriguing multifractal regime.
.

**Materials and Methods**

### Sample preparation and STM/STS characterization

Single-layer NbSe$_2$ was grown by molecular beam epitaxy (MBE) on epitaxial BLG on 6H-SiC(0001) at the HERS endstation of Beamline 10.0.1, Advanced Light Source, Lawrence Berkeley National Laboratory (the MBE chamber had a base pressure of ~ 1 × 10$^{-10}$ Torr). We used SiC wafers with a resistivity of $\rho$ ~ 300 $\Omega$·cm. BLG substrates were obtained by flash annealing SiC(0001) substrates to ~ 1600 K. High purity Nb and Se were evaporated from an electron-beam evaporator and a standard Knudsen cell, respectively. The flux ratio of Nb to Se was controlled to be ~ 1:30. The growth process was monitored by in-situ RHEED and the growth rate was ~17 minutes per layer. During the growth, the substrate temperature was kept at 600 K, and after growth the sample was annealed to 670 K. To protect the film from contamination and oxidation during transport through air to the UHV-STM chamber, a Se capping layer with a thickness of ~10 nm was deposited on the sample surface after growth. For subsequent STM experiments, the Se capping layer was removed by annealing the sample to ~ 530 K in the UHV STM system for 30 minutes. STM imaging and STS experiments were performed in a commercial SPECS GmbH Low-Temperature (1.1 K) STM, under UHV conditions. STM differential conductance (dI/dV) spectra were measured at 1.1 K using standard lock-in techniques. STM/STS data were analyzed and rendered using WSxM software.

### Mean free path estimation (*l*)

The mean free path, *l*, in 2D is $l = h \cdot \sigma /(e^2 \cdot \sqrt{(2\pi n)})$, with $\sigma$ and n the conductivity and electronic density of single-layer NbSe$_2$, respectively. Our previous transport experiments in this type of samples showed $\sigma \approx 1/300$ $\Omega$ in the normal state, therefore $l = 24.4/\sqrt{(n)}$. Taking n = 1.25·10$^{16}$ cm$^{-2}$ as in ref. (*27*), then $l \approx 2$ nm. This estimation may vary due to other factors such as the presence of graphene, the exact geometry of the transport devices, and the exact value of n. In ref. (*27*) a smaller value of $l = 1.3$ nm is reported.

A value of *l* ~ 2 nm yields $\gamma = 4/k_F \cdot l = 4/(\sqrt{(2\pi n)} \cdot l)$ ~ 0.07 << 1, which is in good agreement with the experimental $\gamma$ values extracted from our measurements. Furthermore, the condition $\gamma$ ~ 0.07 << 1 is fulfilled, which validates the applicability of eq.1 used in the theoretical fit of the statistical distributions of the SC width maps.

**Acknowledgments**

We thank Claudia Ojeda-Aristizábal and Reyes Calvo for fruitful discussions. A.M.G.G. thanks Prof. Xi Chen for illuminating discussion and for sharing unpublished work before publication. M.M.U acknowledges support by the Spanish MINECO under grant no. MAT2017-88377-C2-1-R and by the ERC Starting grant "LINKSPM" (Grant No. 758558). This research used resources of the Advanced Light Source, which is a DOE Office of Science User Facility under contract no. DE-AC02-05CH11231. A.M.G.G. acknowledges additional support from a Shanghai talent program and from the National Natural Science Foundation of China (NSFC) (Grant number 11874259).

**Figures and Tables**

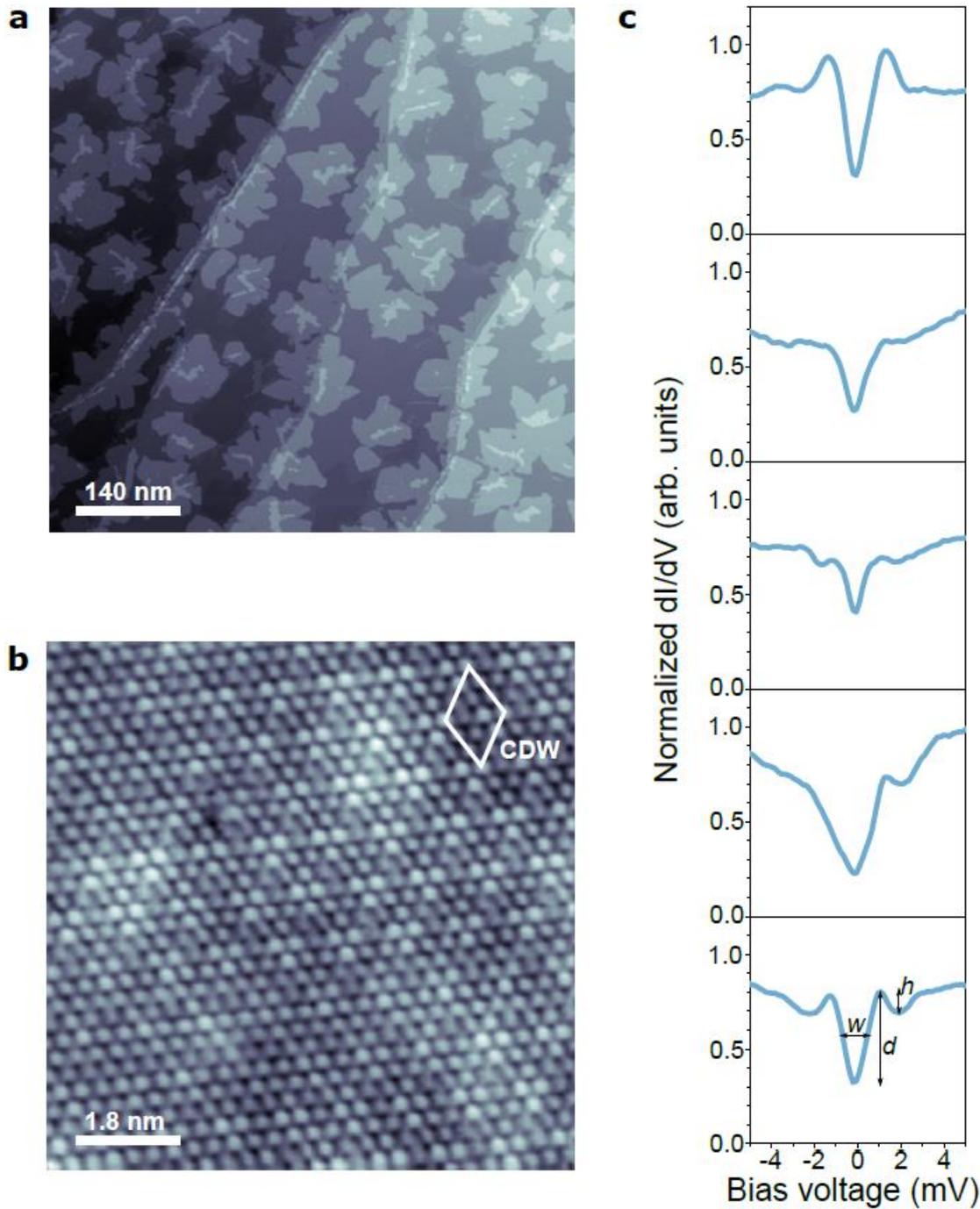

**Figure 1. Fluctuations of the superconducting gap in single-layer NbSe$_2$**. **a**, Large-scale STM topograph of single-layer NbSe$_2$/BLG (V$_s$ =1 V, I$_t$ = 10 pA). **b**, Atomically resolved STM image of single-layer NbSe$_2$. The 3x3 CDW superlattice is indicated (V$_s$ =14 mV, I$_t$ = 1 nA). **c**, Normalized dI/dV spectra acquired at several nearby locations on NbSe$_2$ (f = 938 Hz, I$_t$ = 0.8 nA, V$_{rms.}$ =20 µV). The definition of superconducting gap width, depth and coherence peaks amplitude is indicated with arrows.

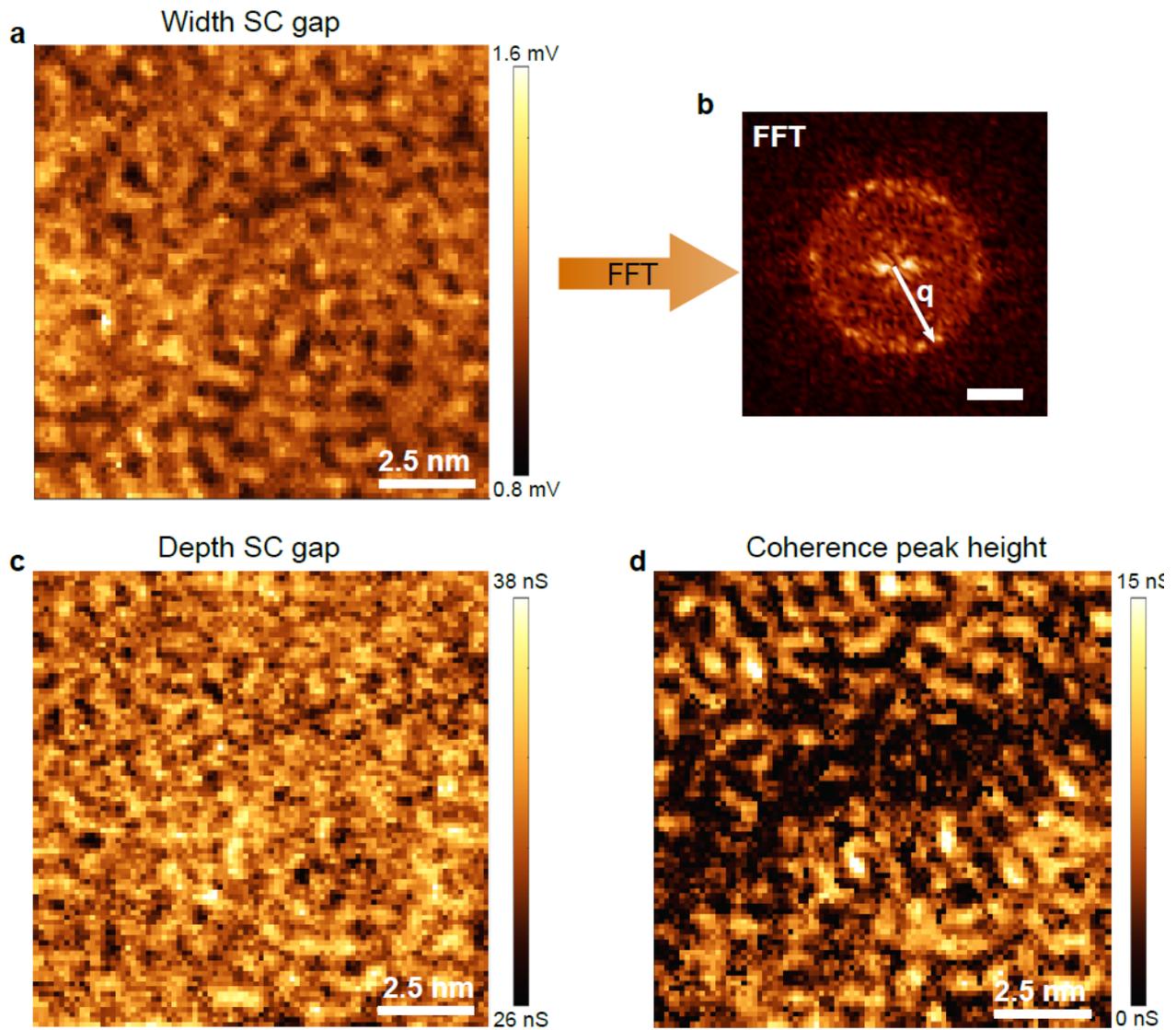

**Figure 2. Real-space fluctuations of the superconducting gap.** Spatial distribution of the superconducting gap width (**a**), depth (**c**) and coherence hole-peak amplitude (**d**) acquired on the same region. **b**, FFT of the gap size distribution in **a**. The observed **q** vector corresponds to a wavelength of 7 Å. Scale bar is 0.5 Å$^{-1}$.

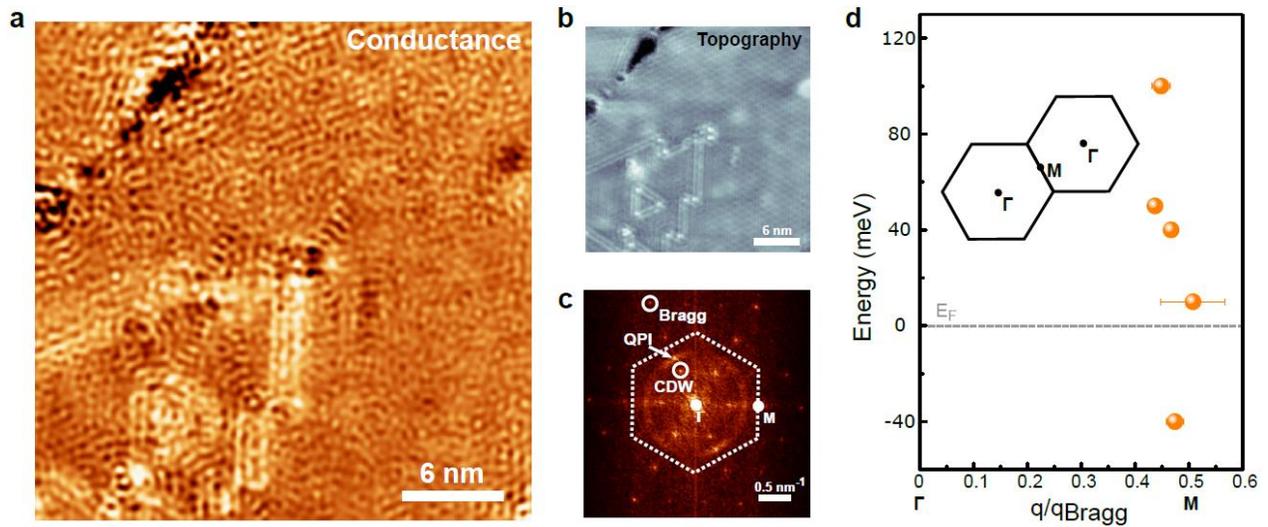

**Figure 3. Quasiparticle interference pattern in single-layer NbSe$_2$. a,** dI/dV conductance map taken at $V_s$ = 40 mV. **b,** Corresponding STM topograph ($V_s$ = 40 mV, $I_t$ = 1 nA), showing the main source of intrinsic defects, i.e., 1D grain boundaries and edges. **c,** FFT of the conductance map in **a**. **d,** Energy dependence of the QPI wavevector along the Γ-M direction extracted from the FFT of the dI/dV maps. Inset: The first/second Brillouin zones.

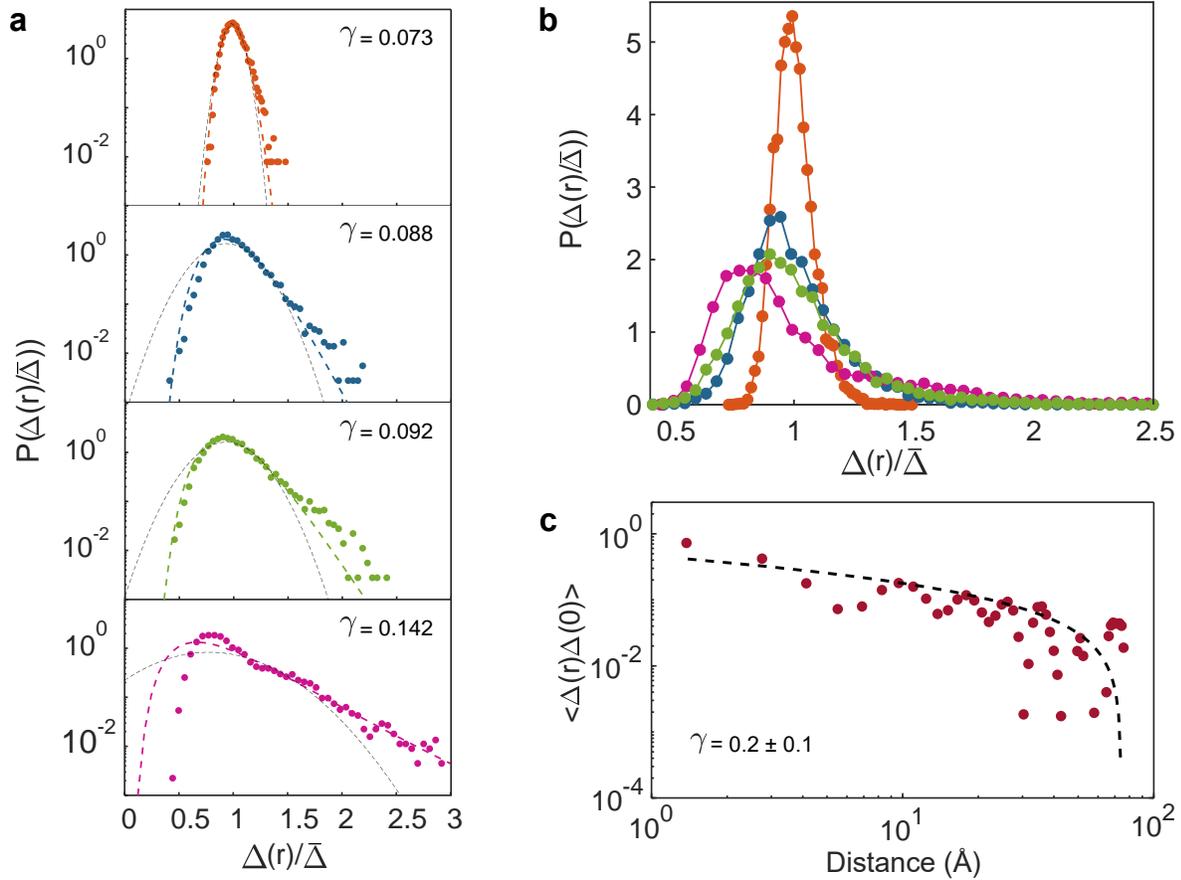

**Figure 4. Multifractal origin of real-space superconducting oscillations. a**, Log-scale SC gap width probability distributions for four different studied regions of single-layer NbSe$_2$ with different disorder strengths ($\gamma = 1/g$, with g the Thouless conductance), normalized to the mean value of the gap. The deviation from a Gaussian distribution (grey dashed curve) and fit to a log-normal distribution from our theoretical model (colored dashed line) indicates the multifractal character. The fitted $\gamma$ value of each distribution is shown in the corresponding upper right panel. **b**, Same distributions merged and shown in linear scale. **c,** Two-point correlation function of the spatial SC gap width map (Fig. 2a) fitted to power-law decay.

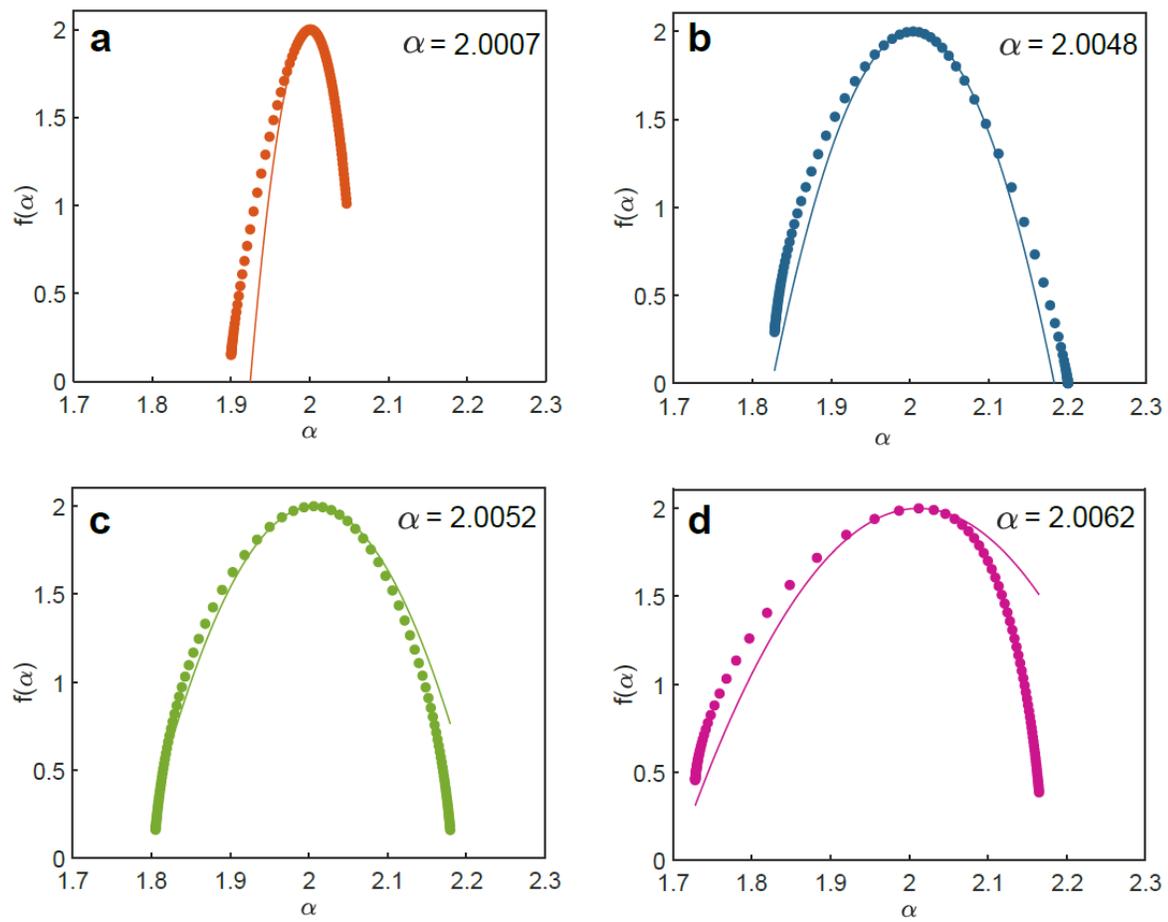

**Figure 5. Singularity spectra. a-d**, Singularity spectra computed from the same experimental spatial distributions Δ(r) studied in figure 4 (dots). Straight lines are the corresponding fits using the analytical prediction for a weakly disordered quasi-2D disordered system.

**Supplementary materials**

### Estimation of the superconducting gap

The fit of the superconducting gap to the prototypical Dynes' formula was elusive due to the spatial fluctuations of the coherence peaks and its absence in many positions. Instead, the superconducting gap parameters were obtained by a reversed peak function, extracting its width and depth.

First, the dI/dV curves are selected in an energy range such as the charge density wave gaps (23) is excluded ($V_S = \pm 2.7$ mV, see shaded area in Fig. S1). Then, the algorithm obtains the minimum and the maximum values of the conductance that define the reversed peak function. The superconducting gap depth is calculated as the difference in conductance between these two points, and the width as the full-width-half-maximum of the reversed peak.

The coherence peaks height is obtained in a similar way. First, positive and negative energy ranges from the shaded area in Fig. S1 are separated. Then, we obtain the energy position of the center of the peak. We define a second point by adding up the energy of the center of the peak plus the full-width-half-maximum of the peak. The difference in conductance of the experimental data between these two points define the coherence peak height.

In Figure S1 we show a prototypical example of one dI/dV curve and the extracted values for the gap width, depth and coherence peak height.

Regarding the dI/dV curves that show no coherence peaks, we are aware that a slight deviation of the real width and depth values may be encountered. Nevertheless, these deviations do not change the real-space modulations of the superconducting gap width nor the asymmetry in the histograms since they represent only the 8% of the dI/dV curves.

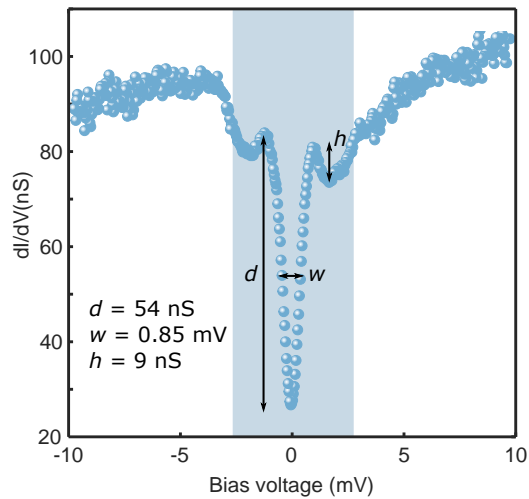

**Figure S1.** Example of width and depth values extracted from the superconducting gap ($V_S = 10$ mV, $I_t = 1$ nA, $V_{rms} = 20$ μV).

## Gap distribution for different areas

In Figure S2 we show the distribution of the superconducting gap width (fig. S2a-d), depth (fig. S2e-h), coherence peaks height (fig. S2 i-l) and conductance at $V_s = 0$ mV (fig. S2m-p) for four different regions of single-layer NbSe$_2$ that show different disorder strengths (from $\gamma = 1/g = 0.073$ to $\gamma = 0.142$). While the associated wave vector for the oscillations remains constant ($\lambda = 7 \pm 1$ Å), both the real-space pattern and the statistical distribution slightly change for different areas.

The asymmetry in the distribution of the superconducting gap width becomes more pronounced as the disorder strength increases, leading to a lower value for the mean gap width and a stronger right-skewed behavior in the width distribution (see Fig. S2a-d), as expected for multifractal eigenstates in the presence of weak disorder.

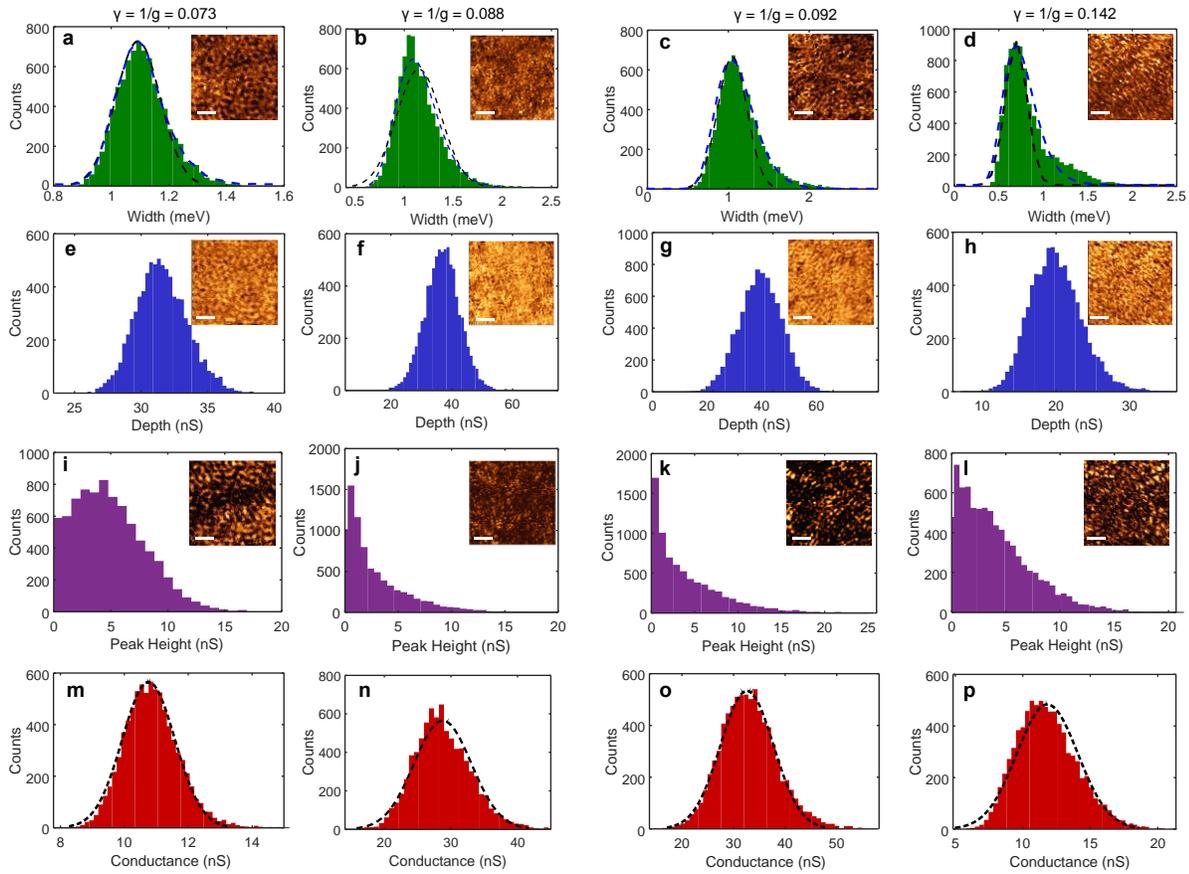

**Figure S2. a-d.** Distribution of the SC gap width on different areas of single-layer NbSe$_2$ and fit of a log-normal distribution (blue dashed line). A Gaussian fit is shown in black dashed line as a reference. The real-space distribution of the gap width in shown in the insets. The scalebar represents a. 2.5 nm, b. 3.6 nm, c-d. 2.5 nm. **e-h.** Distribution of the SC gap depth for the same areas. The real-space distribution is shown in the insets. **i-l.** Distribution of the coherence peak height on the same areas as a-d, respectively. The real-space distribution of the coherence peak height is shown in the insets. **m-p.** Distribution of the conductance at $V_s = 0$ mV for different areas. The dashed line represents the fit to a Gaussian distribution.

In Fig. S2m-p we show the distribution of the conductance values at $V_S = 0$ mV over the same regions of single-layer NbSe$_2$. As discussed in the main text, the distribution is expected to be symmetric since the non-trivial matrix elements induce a different coupling in each point of space. Since the superconducting gap depends exponentially on the coupling, small changes in the coupling will induce large changes in the superconducting gap. Therefore, multifractal effects are amplified and observable through the superconducting gap, while the distribution of the conductance remains symmetric at this level of weak-disorder in the weak-multifractality regime.

For comparison, we show in Fig. S2e-h the distribution of the superconducting gap depth obtained with the same method as the width. Real-space oscillations are also visible as shown in the insets, reflecting the inhomogeneity of the superconducting gap overall shape. Nevertheless, the statistical distribution of the depth is symmetric regardless of the disorder strength. So far, no analytic models attempt to describe the behavior of the depth of the superconducting gap.

The distributions of the coherence peaks height shown in Fig. S2i,l are representative of a type that exhibits the maximum at finite heights, i.e., most of the coherence peaks are present and, therefore, the superconducting phase coherence persists. A second type of distributions (Fig. S2j,l) shows a Poisson shape with a clear maximum at zero height, which indicates much stronger phase fluctuations in those regions. We attribute this distinct behavior to variations in the density of defects among the studied regions, which is inversely proportional to the Thouless conductance (g).

**Relevant spatial length scales**

It is commonly stated in the literature that spatial variations of the order parameter for distances smaller than the superconducting coherence length are strongly suppressed. We show below that in general this is not the case.

The argument for the suppression is largely based on the effective Ginzburg-Landau approach that predicts an essentially constant order parameter for distances shorter than the coherence length. As shown by Gorkov, the Ginzburg-Landau approach agrees with BCS theory but only close to the critical temperature and in general in the range of temperatures in which the experiments were carried out. This is especially true regarding the spatial dependence. Ginzburg-Landau equations result from a gradient expansion of the order parameter where only the leading term, whose typical variation is the coherence length, is kept. The neglected subleading terms, which lead to variations of the order parameter for scales smaller than the coherence length, become relevant when the spatial perturbation, disorder in our case, becomes strong enough. This is certainly the case in case of systems close to a metal-insulator transition. Indeed, there is overwhelming experimental and theoretical evidence of spatial structure of the order parameter for length scales much shorter that the coherence length: in the context of nanograins (*29*, *39–42*) and thin films (*9–11*, *43*). Here it was shown that nanoscale variations of the size of the system induce substantial variations of the spectroscopic gap despite the fact that the coherence length of the materials involved was of the order of hundreds of nanometers. Similarly, recent STM experimental results in ultrathin films disordered conductors have found nanoscale variations of the superconducting in materials where the superconducting coherence length is much longer(*9–11*, *43*).

It is also worth to point out that in disordered superconductors, even neglecting quantum coherent effects, the coherence is typically shorter as is given by the so called "dirty" expression $\xi_D \sim \sqrt{\xi_B l}$ where $\xi_B$ is the bulk clean coherence length and l is the mean free path. Therefore, at least for conventional superconductors, $\xi_D \ll \xi_B$, so we expect variations of order parameter at the nanoscale even without taking into account coherence effects such as localization which are relevant in our system.

In summary, strong disorder reduces drastically the length scale for spatial variations of the order parameter. For weak disorder, and neglecting coherence effects, the length scale is giving by the "dirty" expression which is much shorter than the bulk coherence length. For stronger spatial variations, either because of stronger disorder or because coherent effects, leading for instance to multifractality, the length scale for spatial variations of the order parameter will be even shorter, of the order of the mean free path.

Regarding the mean free path, epitaxial $NbSe_2$ samples are comparable in terms of morphology to those grown by chemical vapor deposition (CVD), on which direct measurement of the mean free path and the electron density has been carried out (*27*). Here, the values reported are $l$ = 1.3 nm. As shown in the main manuscript, our own estimation taking conductivity values obtained in this same type of samples yields a close value of $l \approx$ 2 nm. These values are of the same order of magnitude of the observed SC fluctuations and lie in the lower limit of the spatial correlations of the SC gap (0.7 nm < |r-r′| < 7 nm).

**Wavelength of the superconducting gap modulation**

The modulation of the superconducting gap has an associated wavelength (7Å) that matches the QPI modulation. As discussed in the main text, we propose two plausible mechanisms for the observed correlation between the QPI and the SC modulations: (i) The soft acoustic phonons along the ΓM direction that trigger the formation of the QPI also mediate the formation of Cooper pairs or (ii) a direct spatial correlation between the SC gap fluctuations and the local density of states (LDOS) modulations – QPI pattern –, i.e., Δ(r) ∝ LDOS. In order to gain knowledge here, we have spatially mapped the LDOS in several regions around $E_F$ (± 10 meV) and compare it with the corresponding spatial pattern of the SC order parameter (SC gap width). Surprisingly, the conductance is highly dynamic and the QPI pattern shows rapid spatial variations throughout this energy range (see selected frames in Fig. S3a and the entire movie included in the supplementary material). Importantly, none of the LDOS(E) maps (or QPI patterns) shows significant correlation (or anticorrelation) with the corresponding pattern of the SC order parameter in that region (see Fig. S3b) In other words, the spatial fluctuations of the SC order parameter do not significantly follow (ε ≈ ± 1) the modulations of the DOS for any energy within ± 10 meV. Therefore, we can rule out the second scenario in which Δ(r) ∝ LDOS. The formation of the Cooper pairs mediated by the same soft acoustic phonons triggering the QPI in this system remains compatible with our observations.

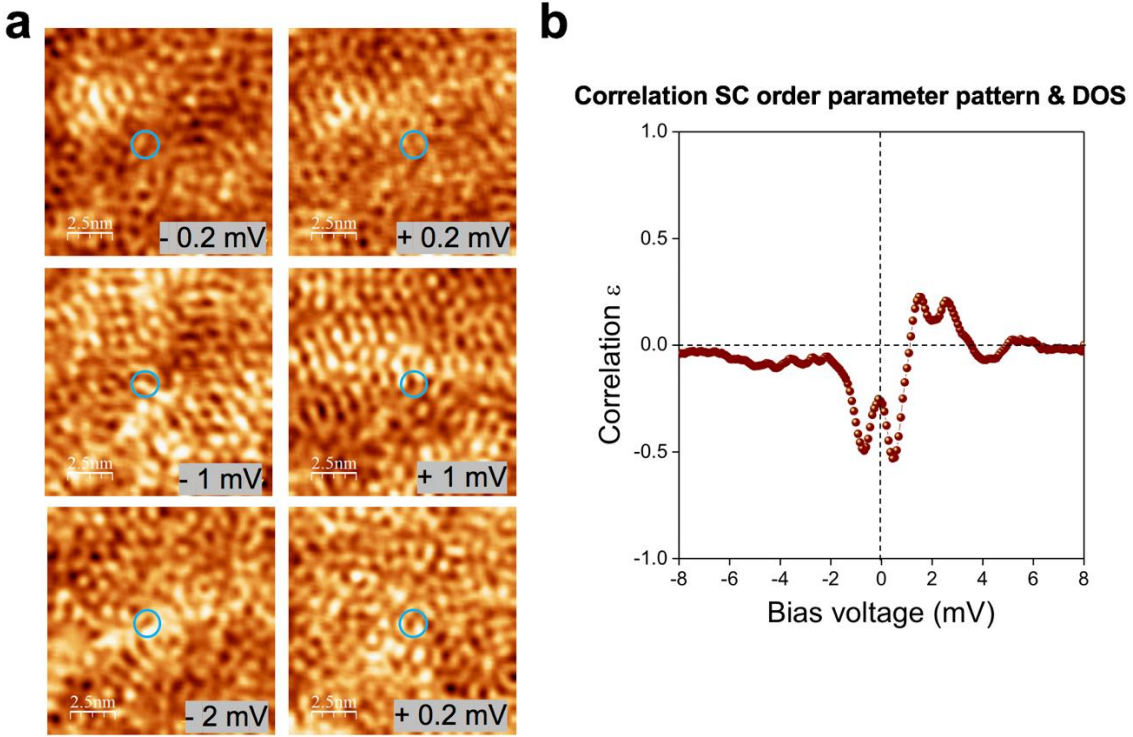

**Figure S3. a.** dI/dV conductance maps simultaneously obtained at ± 0.2, ± 1.0 and ± 2.0 mV. The blue circle indicates the same region to highlight the DOS changes. **b.** 2D correlation between the dI/dV conductance map and the spatial distribution of the SC gap width as a function of the bias voltage.

## Inhomogeneous BCS superconductivity

We now provide a brief summary of the calculation leading to a log-normal distribution function of the order parameter amplitude and the two-point correlation function employed in the main text. We follow Ref. (*20*) closely and refer to it for further details.

The starting point of our analysis is the Bogouliubov-de Gennes(BdG) Hamiltonian,

$$H = \int d\mathbf{r} \left[ \sum_\sigma \Psi_\sigma^\dagger(\mathbf{r}) \left( -\frac{\hbar^2}{2m}\nabla^2 + U(\mathbf{r}) - \mu \right) \Psi_\sigma(\mathbf{r}) \right.$$
$$\left. + \Delta(\mathbf{r})\Psi_\downarrow^\dagger(\mathbf{r})\Psi_\uparrow^\dagger(\mathbf{r}) + \text{h.c.} \right] \tag{S1}$$

where $\psi_\sigma^\dagger(r)$ creates an electron in position eigenstate **r** and spin σ U(r) is a random potential,

$$\Delta(\mathbf{r}) = -\frac{\lambda}{\nu(0)} \langle \Psi_\uparrow(\mathbf{r})\Psi_\downarrow(\mathbf{r}) \rangle, \tag{S2}$$

**λ** is the dimensionless BCS coupling constant and ν(0) is the bulk density of states at the Fermi energy. Assuming that the spatial part of $\psi_\sigma^\dagger(r)$ is proportional to the eigenstates of the one-body problem $\psi_n(r)$, an approximation which should be valid in the limit of weak coupling and disorder, it is possible to derive a modified BCS gap equations,

$$\Delta(\epsilon) = \frac{\lambda}{2} \int_{-\epsilon_D}^{\epsilon_D} \frac{I(\epsilon, \epsilon') \Delta(\epsilon')}{\sqrt{\epsilon'^2 + \Delta^2(\epsilon')}} d\epsilon' \tag{S3}$$

where $\varepsilon_D$ is the Debye energy and $\Delta(\varepsilon)$ is the superconducting gap as a function of energy. The BCS interaction matrix elements are $I(\varepsilon, \varepsilon') = V \int_{-\varepsilon_D}^{\varepsilon_D} dr |\psi(\varepsilon, r)|^2 |\psi(\varepsilon', r)|^2$, where $\psi_n(\varepsilon, r)$ is the eigenstate of the one-body problem of energy $\varepsilon$.

As it was mentioned previously, eigenfunctions of a disordered system close to a metal-insulator transition are multifractal (*37*). There are different ways to characterize this multifractality, for instance in the inverse participation ratio (IPR)(*2, 35*),

$$P_q = \int d\mathbf{r} |\psi(\mathbf{r})|^{2q} \sim L^{d_q(q-1)}, \tag{S4}$$

is related to an anomalous scaling with the system size. A signature of multifractality in eigenstate correlation functions,

$$I(\epsilon, \epsilon') = \left(\frac{E_0}{|\epsilon - \epsilon'|}\right)^\gamma \tag{S5}$$

is a slow power-law decay with $\gamma = 1 - \frac{d_2}{d}$ which is cutoff for energy differences smaller than the mean level spacing and larger than $E_0 \sim (\nu(0) L_0^2)^{-1}$ an energy scale related to the minimum length $L_0$ for which fractal behavior is observed. Typically, $L_0$ is of the order of the mean free path, l. For weak disorder in quasi-two dimensions, $L_0 \sim l \sim g/k_F$ and therefore $L_0 \sim l \sim g/k_F$ which we expect to be an energy scale between the Fermi and the Debye energy. The exponent $\gamma$ is directly related to the disorder strength, $\gamma \approx \frac{1}{g} \ll 1$ where $g \gg 1$ is the dimensionless conductance.

We have now all the information necessary to summarize the calculation of the distribution function of the order parameter. In the limit $\gamma \ll 1$ it is possible to compute $\Delta(\varepsilon)$ analytically by expanding $I(\varepsilon, \varepsilon')$ in powers of $\gamma$ and solving the gap equation Eq. S3 order by order. The spatial dependence of the order parameter amplitude is given by,

$$\langle \Delta(\mathbf{r}) \rangle = \frac{\lambda V}{2} \int \frac{\Delta(\epsilon)}{\sqrt{\Delta(\epsilon)^2 + \epsilon^2}} |\psi(\epsilon, \mathbf{r})|^2 d\epsilon, \tag{S6}$$

and higher moments,

$$\langle \Delta^n(\mathbf{r}) \rangle = \int d\mathbf{r} \prod_{j=1}^{n} \left( \frac{\lambda V}{2} \int \frac{\Delta(\epsilon_j)}{\sqrt{\Delta(\epsilon_j)^2 + \epsilon_j^2}} |\psi(\epsilon_j, \mathbf{r})|^2 d\epsilon_j \right), \tag{S7}$$

are evaluated analytically by a similar expansion in $\gamma \ll 1$, keeping only the leading terms. The final result is,

$$\langle \Delta^n(\mathbf{r}) \rangle \propto e^{\gamma \ln(\epsilon_D/E_0)(3n-n^2)/2}. \tag{S8}$$

A similar expression was derived in Ref. (*18*) in the limit of strong disorder $g \sim 1$ where the perturbative expansion in $g$ should break down.

These are the moments of a log-normal distribution,

$$\mathcal{P}\left(\frac{\Delta(\mathbf{r})}{\bar{\Delta}}\right) = \frac{\bar{\Delta}}{\Delta(\mathbf{r})\sqrt{2\pi}\sigma} \exp\left[-\frac{\left(\ln\left(\frac{\Delta(\mathbf{r})}{\bar{\Delta}}\right) - \mu\right)^2}{2\sigma^2}\right] \tag{S9}$$

wit h $\mu = \gamma \ln\left(\frac{\varepsilon_D}{E_0}\right)/2$, $\sigma = \sqrt{\gamma \ln\left(\frac{E_0}{\varepsilon_D}\right)}$

$$\bar{\Delta} = D(\gamma)\epsilon_D \left(1 + \frac{\gamma}{\lambda}\left(\frac{\epsilon_D}{E_0}\right)^\gamma\right)^{-\frac{1}{\gamma}} \left(\frac{\epsilon_D}{E_0}\right)^\gamma, \tag{S10}$$

and

$$D(\gamma) = \left(\frac{\gamma \Gamma(\frac{1}{2}(1-\gamma))\Gamma(\frac{\gamma}{2})}{2\sqrt{\pi}}\right)^{\frac{1}{\gamma}}. \tag{S11}$$

Finally, we compute the two-point correlation function $\langle \Delta(r)\Delta(r') \rangle$. For multifractal eigenstates[S15]

$$\langle |\psi(\epsilon, \mathbf{r})|^2 |\psi(\epsilon', \mathbf{r'})|^2 \rangle \sim \frac{1}{|r-r'|^\gamma |\epsilon - \epsilon'|^{\gamma/d}} \tag{S12}$$

where the decay in energy is cutoff in the way explained previously. Regarding the power-law decay in space, multifractality cannot be observed for distances shorter than the mean free path. For finite size systems, multifractality is also restricted to distances shorter than the system size. Similarly, for systems close but not at the critical point $\varepsilon_c$, multifractality is restricted to lengths smaller than $\xi \sim |\varepsilon - \varepsilon_c|^{-\nu}$ even in the thermodynamic limit. Finally, we note that, for eigenstates separated in energy by $\omega = \varepsilon - \varepsilon'$, the spatial power-law decay cannot hold for distances longer than $L_\omega \sim (\frac{1}{\omega\rho})^{1/d}$ (*37*). The decay for distances larger than $\zeta, L_\omega$ is not universal and likely exponential in most cases. From Eq. S6, it is straightforward to show that this power-law decay of eigenstate correlations translates into a similar power-law decay of the order parameter spatial correlations for intermediate distances $l < |r - r'| < L_\omega, \zeta$:

$$\langle \Delta(r)\Delta(r') \rangle \propto \frac{1}{|r-r'|^\gamma}. \tag{S13}$$

In the main text we compare these analytical predictions with experimental results in different zones of the sample. Overall, we have found good agreement between theory and experiment in all zones for both the gap distribution function and the two-point correlation function. Each zone is described by a different $g \gg 1$, which is typical of two or quasi two-dimensional superconductors where features expected at the Anderson metal-insulator transition are observed for a broad range of values of the dimensionless conductance provided that the localization length is larger than the system size.

**Singularity spectrum *f(α)***

### 1. Method to compute the experimental singularity spectrum *f(α)*

The singularity spectrum $f(\alpha)$ provides information about the scaling properties of a given probability measure $|P(r_i)|^2$. Among others, it has been employed to characterize multifractality in the context of turbulence and the Anderson metal-insulator transition(*35, 38, 44*). Following the method introduced in Ref.(*44*), we provide a brief introduction about the calculation method of the singularity spectrum. We apply it to a normalized experimental and theoretical order parameter $\Delta(r_i)$, $|P(r_i)|^2 = \frac{\Delta(r_i)}{\sum_j \Delta(r_j)}$.

In order to proceed, we define $\tau_q, \alpha_q$ and $f(q)$ as follows:

$$\sum_i |P(r_i)|^{2q} \sim \left(\frac{1}{N}\right)^{\tau_q}$$

where N is the number of small boxes of typical size 1/N in which we divide the volume of the system.

The exponent $\tau_q$ therefore can be written as, $\tau_q = -\frac{\ln \sum_i |P(r_i)|^{2q}}{\ln N}$.

Similarly, $\alpha_q$ and $f(q)$ are given by:

$$\alpha_q = -\frac{1}{\ln N} \sum_{i=1}^{N} \frac{|P(r_i)|^{2q}}{\sum_{j=1}^{N}|P(r_j)|^{2q}} \ln \frac{|P(r_i)|^2}{\sum_{j=1}^{N}|P(r_j)|^2} = -\frac{1}{\ln N} \frac{\sum_{i=1}^{N}|P(r_i)|^{2q} \ln|P(r_i)|^2}{\sum_{j=1}^{N}|P(r_j)|^{2q}}$$

$$f(q) = -\frac{1}{\ln N} \sum_{i=1}^{N} \frac{|P(r_i)|^{2q}}{\sum_{j=1}^{N}|P(r_j)|^{2q}} \ln \frac{|P(r_i)|^{2q}}{\sum_{j=1}^{N}|P(r_j)|^{2q}}$$

$$= -\frac{1}{\ln N} \frac{q \sum_{i=1}^{N}|P(r_i)|^{2q} \ln |P(r_i)|^2}{\sum_{j=1}^{N}|P(r_j)|^{2q}} + \frac{1}{\ln N} \frac{\sum_{i=1}^{N}|P(r_i)|^{2q} \ln \sum_{j=1}^{N}|P(r_j)|^{2q}}{\sum_{j=1}^{N}|P(r_j)|^{2q}}$$

$$= -\frac{1}{\ln N} \frac{q \sum_{i=1}^{N}|P(r_i)|^{2q} \ln |P(r_i)|^2}{\sum_{j=1}^{N}|P(r_j)|^{2q}} + \frac{\ln \sum_{j=1}^{N}|P(r_j)|^{2q}}{\ln N}$$

From the expressions for $\tau_q$, $\alpha_q$ and $f(q)$, we get the singularity spectrum $f(\alpha_q)$ associated to $|P(r_i)|^2$ by using the Legendre transformation,

$$f(\alpha_q) = q\alpha_q - \tau_q$$

where $\alpha_q = \frac{d\tau_q}{dq}$.

In the case of the Anderson transition in $2 + \epsilon$ dimensions, or in two dimensions for sizes smaller than the localization length, that occur in the weak-disorder, weak-multifractal limit, the probability measure is just the density of probability associated to a given eigenstate of the Schrödinger equation. The singularity spectrum is exactly parabolic $f(\alpha_q) = d - \frac{(\alpha_q - \alpha_0)^2}{4(\alpha_0 - d)}$ (refs.(35, 38)), where, $\alpha_0 = d + \epsilon$, $d = 2$ is the dimension of the system, and $\epsilon \approx \gamma$.

2. **Numerical calculation of the *f(α)* spectrum using Bogoulibov de Gennes (BdG) equations**

As it was shown in the main manuscript, the experimental *f(α)* spectra (figure 6) show some deviations from the parabolic prediction. Here we compare it with the theoretical expectation by computing numerically $|P(r_i)|^2$ the two-dimensional Bogoliubov de Gennes equation in a random potential with a box distribution and then the associated singularity spectrum. As is observed in Fig S4., there is good qualitative agreement with the experimental results. Deviations from the parabolic prediction are similar to those found in the comparison with the experimental results. We stress that these deviations are expected. In the case of superconductivity, the measure of probability is, within BCS, a weighted sum about eigenstates of the one-body problem around the Fermi energy, while the parabolic prediction applies only to single eigenstates of the one-body problem with energies very close to the mobility edge where the metal insulator transition takes place.

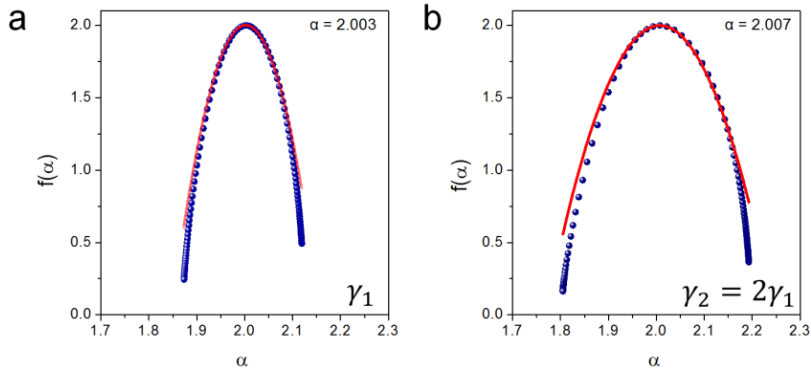

**Figure S4:** Singularity spectrum *f(α)* of Δ(r) in the Bogoliugov de Gennes formalism. Blue circles correspond to the numerical solution of the two-dimensional BdG equations in a 80 x 80 lattice with a random onsite potential with increasing disorder ($\gamma_1$ and $\gamma_2 = 2\gamma_1$), $E_D = 0.15 \cdot E_F$ and the electron-phonon coupling $\lambda = 1$. We find qualitative same behavior in the experimental results as in the parabolic prediction in the limit of weak multifractality (red dotted line).